\documentclass[letterpaper]{emulateapj}                %
\usepackage{epsfig,graphicx,latexsym,amsmath,amssymb}                         %
\usepackage{natbib}                                                           %
\bibpunct[,]{(}{)}{;}{a}{}{,}                                                 %
\pdfoutput=1

\def\pderiv(#1/#2){\frac{\partial#1}{\partial#2}}                   %

\begin{document}

\author{Miguel Preto\altaffilmark{1,2},
Ingo Berentzen\altaffilmark{1,3}, Peter Berczik\altaffilmark{4,5,1} 
\& Rainer Spurzem\altaffilmark{4,1,6}}
\altaffiltext{1}
{Astronomisches Rechen-Institut, Zentrum f\"{u}r Astronomie, University of Heidelberg, 
D-69120 Heidelberg, Germany}
\altaffiltext{2}
{Max Planck Institut f\"ur Gravitationsphysik
(Albert-Einstein-Institut), D-14476 Potsdam, Germany}
\altaffiltext{3}
{Institut f\"ur Theoretische Astrophysik, Zentrum f\"ur Astronomie der University 
of Heidelberg, Albert-Ueberle-Str. 2, D-69120 Heidelberg, Germany}
\altaffiltext{4}
{National Astronomical Observatories of China, Chinese Academy of Sciences NAOC/CAS, 
20 A Datun Rd., Chaoyang District, Beijing 100012, China}
\altaffiltext{5}
{Main Astronomical Observatory (MAO), National Academy of Sciences of Ukraine (NASU), 
Akademika Zabolotnoho 27, 03680 Kyiv, Ukraine)}
\altaffiltext{6}
{The Kavli Institute for Astronomy and Astrophysics at Peking University}
%\date{\today}

\title{Fast coalescence of Massive Black Hole Binaries from Mergers of Galactic Nuclei: 
Implications for Low-Frequency Gravitational-Wave Astrophysics}

\begin{abstract}
We investigate a purely stellar dynamical solution to the Final Parsec Problem. Galactic nuclei 
resulting from major mergers are not spherical, but show some degree of triaxiality. With $N$-body
simulations, we show that massive black hole binaries (MBHB) hosted by them will continuously 
interact with stars on centrophilic orbits and will thus inspiral---in much less than a Hubble 
time---down to separations at which gravitational wave (GW) emission is strong enough to drive 
them to coalescence. Such coalescences will be important sources of GWs for future space-borne 
detectors such as the {\it Laser Interferometer Space Antenna} (LISA). Based on our results, we 
expect that LISA will see between $\sim 10$ to $\sim {\rm few} \times 10^2$ such events every year, 
depending on the particular MBH seed model as obtained in recent studies of merger trees of
galaxy and MBH co-evolution. Orbital eccentricities in the LISA band will be clearly distinguishable 
from zero with $e \gtrsim 0.001-0.01$. 
\end{abstract}


\section{Introduction}
Massive black hole binaries (MBHBs) are one of the most interesting sources of gravitational 
waves (GWs) for future space-borne detectors such as the {\it Laser Interferometer Space Antenna} 
(LISA). They are expected to coalesce under the strong emission of GWs, after stellar- and/or 
gas-dynamical processes bring them to separations small enough ($a_{GW} \sim 10^{-3}$ pc) that 
GW emission is efficient in making them coalesce in less than a Hubble time 
\citep{2003ApJ...596..860M,2005ApJ...634..921A}. It is still an open problem whether MBHB 
coalescences are generic and prompt, or whether long-lived binaries are the norm. 

The paradigm for MBH binary evolution, after a merger of gas-poor galaxies, consists of 
three distinct phases \citep{1980Natur.287..307B}. First, the two MBHs sink towards the 
center due to the dynamical friction exerted by the stars. This process continues after 
they form a bound pair at a semimajor axis separation $a \sim r_h$, where $r_h$ is the 
binary's influence radius defined to be the radius which encloses twice the mass of the 
binary in stars. It stops when the binary reaches the {\it hard binary} separation $a 
\sim a_h$ \citep{1996NewA....1...35Q,2002MNRAS.331..935Y}
\begin{equation}
a_h := \frac{G \mu_r}{4 \sigma^2} \sim \frac{1}{4} \frac{q}{(1+q)^2} r_h,
\label{eqn1}
\end{equation} 
where $\mu_r$ is the binary's reduced mass, $\sigma$ is the local $1$D velocity dispersion,
$q=M_{\bullet,2}/M_{\bullet,1}$ is the binary's mass ratio. Secondly, for $a \lesssim a_h$, as dynamical 
friction becomes inefficient in further driving the inspiral, it is instead the {\it slingshot} 
ejection of stars, following three-body scattering with the binary, that dominates. Thirdly, 
the binary eventually reaches a separation $a_{GW}$ at which the loss of orbital energy to GW 
emission drives the final coalescence. The transition from the first to the second phase 
is prompt provided that the mass ratio of the remnants is not too small $q = M_2/M_1 
\gtrsim 0.1$ \citep{2009arXiv0906.4339C,2011ApJ...729...85C}. In contrast, the subsequent 
transition from the second to the third phase could constitute a bottleneck for the binary 
evolution towards final coalescence. This is the so-called {\it Final Parsec Problem}.

In {\it quasi-steady spherical} stellar environments, the binary's hardening rate $s(t) \equiv
d/dt(1/a)$ slows down significantly once it reaches separations a few times below $\sim a_h$ 
\citep{1997NewA....2..533Q,2003ApJ...596..860M,2005ApJ...633..680B}. In these spherical and 
gas-poor nuclei, two-body relaxation is the only mechanism for populating the binary's loss 
cone \footnote{The loss cone is the region of phase space corresponding, roughly speaking, 
to orbits that cross the binary, {\it i.e.} with angular momentum $J \lesssim J_{lc} =
\sqrt{GM_{12}fa_{bin}}$, where $f={\mathcal O}(1)$ \citep{1977ApJ...211..244L}.}, but being a 
slow diffusive process, it is only in low-luminosity galaxies harboring MBHs of mass 
$M_\bullet \lesssim {\rm few} \times 10^6 M_\odot$ that central relaxation times are short 
enough to drive the binary to coalescence in less than a Hubble time \citep{2007ApJ...671...53M}.

But spherical models are a {\it worst case scenario}---and not a very realistic one at that! 
Merger remnants will generally be irregular with some degree of triaxiality and, even if triaxiality 
would only be a rather mild and transient phenomenon, it may suffice to bring the binary down to 
$a_{GW}$ \citep{2004ApJ...606.788P}. \cite{2006ApJ...642L..21B} and \cite{2009ApJ...695..455B} 
studied triaxial, rotating models of galactic nuclei using $N$-body simulations---including the 
full post-Newtonian corrections to the MBHB. They have shown that MBHBs in such models do indeed
coalesce in much less than a Hubble time. The next logical step is to study mergers of galactic 
nuclei to investigate if the latter results still hold true under more realistic models and initial 
conditions. 

In this Letter, we use N-body simulations to show that: (1) in 
merging nuclei, the hardening rate is $N$-independent---allowing the extrapolation of $N$-body 
results to real galaxies; (2) the triaxiality depends on the orbital parameters of the progenitor 
galaxies: prolate shapes occur when the merger is almost radial, while an oblate morphology is 
the result of a less radial merger; (3) MBHs become bound with high eccentricities (up to $e 
\sim 0.95$); (4) the eccentricity tends, on average, to increase in good agreement---often 
quantitative---with \cite{1996NewA....1...35Q} predictions; (5) high eccentricities assist the 
MBHB into promptly coalescencing in much less than a Hubble time; (6) eccentricities in the 
LISA band are likely to be distinguishable from zero ($e \gtrsim 0.001-0.01$) even though GW 
circularizes the orbits, and will also be quite large ($0.4 \lesssim e \lesssim 0.8$) in the 
Pulsar Timing Array (PTA) band.

\section{Models and Initial Conditions}
We have performed two sets of $N$-body experiments. In both sets, galactic nuclei are 
represented by spherically symmetric Dehnen models \citep{1993MNRAS.265..250D,1994AJ....107..634T}. 
These models have a central power law density profile, $\rho(r)=(3-\gamma)M_T/4\pi 
r^\gamma(r_b+r)^{4-\gamma}$, with logarithmic slope $\gamma$ and a break radius $r_b$ which are both
set equal to one. The total mass of each nucleus is set $M_T=1$, and we adopt units where $G=1$. 
The total mass of the binary MBH $M_{12}=M_{\bullet,1}+M_{\bullet,2}$, and we take 
$q=M_\bullet,2/M_\bullet,1=1$. We study unequal mass MBH coalescences in parallel papers 
\citep{PBNPS11,PB11}.

The set (A) consists of a single spherical nucleus where two MBHs are placed symmetrically 
about the center, on an unbound orbit, with initial separation $\Delta r_0=2$, initial angular 
momentum $L/L_c=0.5$, where $L_c$ is the angular momentum of the local circular orbit. The set
(B) consists on the equal-mass merger of two initially bound---but well-separated---spherical 
nuclei, each of which has a single MBH at the center with zero initial velocity with respect 
to its nucleus.  For B, the initial separation $\Delta r_0$ refers to both nuclei taken as if 
they were point masses located at each center of mass. The half-mass radius of each nucleus 
is $r_{1/2} \approx 2.41$; accordingly, and in order to have an initial configuration with 
two well separated nuclei, while minimizing the computing time, we set the initial separation 
equal to $20$. For the initial orbital angular momentum of the binary nuclei, we have taken 
two values $L/L_c=0.14$ and $0.6$ given the nearly-parabolic encounters typical of major galaxy 
mergers seen in cosmological simulations \citep{2006A26A...445..403K}. During the first 
pericenter passages, the MBH separations are $\Delta r_{BH} \sim 0.2$ $\sim 0.1 r_{1/2}$ and 
$\Delta r_{BH} \sim 2.2 \sim r_{1/2}$, respectively. Table 1 lists the runs and adopted 
parameters.
%\vspace{0cm}
\begin{table}[h,t]
%\vspace{0cm}
\centering{
\begin{tabular}{c||c|c|c|c|c}
 $M_{12}{\backslash}N$  &    64K  &  128K  &  256K    &  512K  & 1M   \\
\toprule
0.005  &  $\lambda_{sph}$     & $\lambda_{sph}$	   &  $\lambda_{sph}$  &  $\lambda_{sph}$  &   $\lambda_{sph}$   \\
0.1  &  $\lambda_{sph}$     & $\lambda_{sph}$	   &  $\lambda_{sph}$  &  $\lambda_{sph}$  &   ---   \\
\toprule
0.005  &  $\lambda_2$     & $\lambda_2$	   &  $\lambda_2$  &  $\lambda_2$  &   $\lambda_2$   \\
0.01  &  $\lambda_1, \lambda_2$  & $\lambda_1,\lambda_2$    &  $\lambda_1,\lambda_2$  &  $\lambda_1,\lambda_2$  &  ---   \\
0.02  &  $\lambda_1, \lambda_2$  & $\lambda_1,\lambda_2$    &  $\lambda_1,\lambda_2$  &  $\lambda_1,\lambda_2$  &  ---   \\
0.1  &  $\lambda_1, \lambda_2$  & $\lambda_1,\lambda_2$    &  $\lambda_1,\lambda_2$  &  $\lambda_1,\lambda_2$  &  ---   \\
\end{tabular}}
%\vspace{0.cm}
\caption{$N$-body integrations. $1^{\rm{st}}$ column: mass of the MBH binary; Other columns: particle number $N$;
First two lines: simulations of spherical nuclei; Last four lines: Simulations of merging nuclei; $\lambda=L/L_c$ 
measures the initial orbital angular momentum of the MBH binary ($\lambda_{sph}=0.5$ for spherical nuclei), or 
otherwise it measures the initial orbital angular momentum of the merging nuclei ($\lambda_1=0.14$ for 
near-radial merger and $\lambda_2=0.6$ for less radial merger). All nuclei have $\gamma=1$; all binaries
have equal mass $q=M_{\bullet,2}/M_{\bullet,1}=1$.}
\label{tb1}
\end{table}

We have performed the $N$-body simulations using the parallel $\varphi$-GPU code. 
This is a yet unpublished variant of the parallel direct N-body code $\varphi$-GRAPE 
\citep{2007NewA...12..357H}, which uses GPU accelerator cards on parallel clusters 
\citep{BNHS11}. It includes a fourth-order Hermite integration scheme, with block 
time steps, analogous to NBODY1 \citep{2003gnbs.book.....A}.

The code does not include regularization of close encounters, and softening of the gravitational 
interaction is adopted instead. The softening length has to be chosen small enough that it 
reproduces the refilling of the binary's loss cone by two-body relaxation. After some testing, 
we adopt a softening length $\epsilon=10^{-4}$ in model units. We set the time step parameter 
\citep{2003gnbs.book.....A} to $\eta_*=0.01$ for the field stars and $\eta_{BH}=0.001$ for the BHs. 
Furthermore, we force the MBHs to be advanced synchronously at all times with the smallest step. 
With the parallelized version of the $\varphi$-GPU code, one can study models with very large 
number $N$ of particles and the results agree with NBODY4 \citep{2003gnbs.book.....A} as far as 
single stars and distant encounters are concerned. For the high velocity dispersions present in 
nuclei with a MBH, the effect of close encounters between field stars is negligible for the 
{\it bulk} evolution of the stellar system \citep{2010ApJ...708L..42P}.

\section{MBH evolution in spherical versus in merging nuclei}
\begin{figure}
\vspace{-0.4cm}
\epsscale{1.3}
\plotone{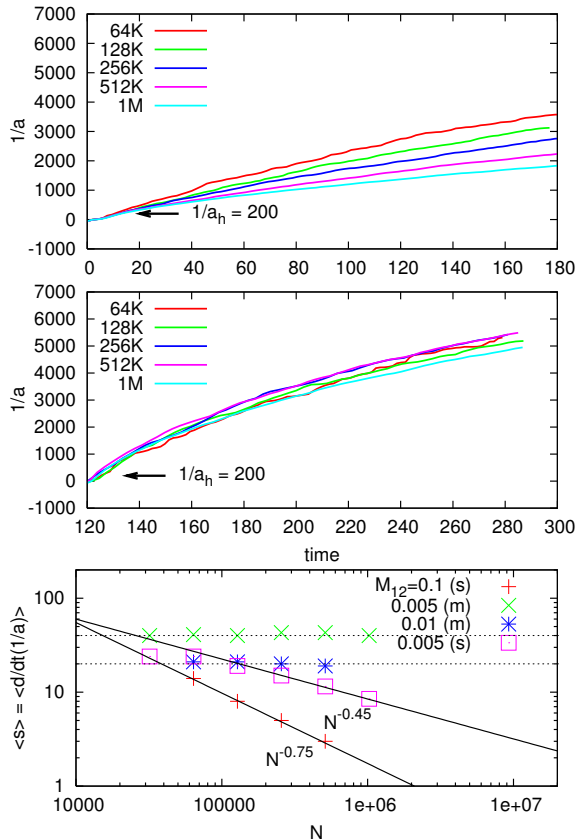}
\vspace{-1cm}
\caption{Binary hardening. Upper panel: in a spherical nucleus, $s(t)$ decreases with $N$. 
Middle panel: in a merging nucleus, $s(t)$ is N-independent. Lower panel: hardening rates 
as a function of $N$ for different $M_{12}$. Being much smaller, $\langle s \rangle$ of the 
$M_{12}=0.1$ binary has been multiplied by $100$ to better fit in the plot. Labels 's' for 
spherical and 'm' for merger.}
\label{fig1}
\end{figure}
The stars that drive the orbital decay of a hard MBHB are those that enter the loss cone
orbits. The MBHB's hardening rate is thus determined by the product of the flux of stars 
entering the loss cone with the average kinetic energy they receive when  ejected---at the 
expense of the MBHB's orbital energy---through the slingshot mechanism. Denoting by 
$\mathcal F_{lc}(E,t)$ the time-dependent flux into the loss cone and by $\langle \Delta E(E) 
\rangle$ the mean kinetic energy imparted to stars which are scattered off by the binary, 
the hardening rate is given by
\citep{2002MNRAS.331..935Y}
\begin{equation}
\frac{d}{dt} \left( \frac{1}{a}  \right) = \frac{2m_*}{GM_{12}\mu_r} 
\int_0^{+\infty} dE  \langle \Delta E(E) \rangle \mathcal F_{lc}(E,t),
\label{eqn2}
\end{equation} 
where $E=GM_{12}/r + \Phi_*(r)-1/2 \ v^2$, and $\Phi_*(r)$ is the gravitational potential due to 
the stars. The mean kinetic energy $\langle \Delta E(E) \rangle$ is given by 
\begin{equation}
\langle \Delta E(E) \rangle \sim \langle C \rangle \frac{G  \mu_r}{a}, 
\label{eqn3}
\end{equation}
where $\langle C \rangle \approx 1.25$ is a dimensionless quantity which was measured from 
three-body scattering experiments \citep{1996NewA....1...35Q}. Therefore, the hardening rate 
$s(t)$ can be rewritten as
\begin{equation}
s(t) \equiv \frac{d}{dt} \left( \frac{1}{a} \right) \approx \frac{2 m_* \langle C \rangle}{M_{12} a}
\int_0^{+\infty} dE \mathcal F_{lc}(E,t).
\label{eqn4}
\end{equation}
The time evolution of the flux $\mathcal F_{lc}(E,t)$ depends on the symmetries of the gravitational
potential---and on the orbit families it supports---of the nuclei in question. In principle, 
$\mathcal 
F_{lc}(E,t)$ in the spherical case can be obtained from Fokker-Planck calculations that take into 
account the diffusion of stars in phase space \citep{2007ApJ...671...53M,2010ApJ...708L..42P}. Here 
we derive simple scaling relations which are useful in interpreting the N-body results. For each 
energy $E$, $\mathcal F_{lc}(E,t) \propto n(E,t)/T_{rlx}(E,t)$ where $n(E,t)$ is the number of stars 
of energy $E$ per unit energy and $T_{rlx}(E,t)$ is the local two-body relaxation time; the latter 
scales as $T_{rlx} \propto \sigma^3/\rho m_*$ \citep{1987degc.book.....S}. The flux of stars 
into the loss cone is expected to peak around $r_h$ \citep{2008ApJ...677.146P}, so we evaluate 
these quantities there. Hence, $\sigma_h^2 \sim G(M(<r_h)+M_{12})/r_h \sim 3 G M_{12}/r_h 
\propto M_{12}^{1/2}$---where $r_h \propto M_\bullet^{1/2}$ follows from the $M_\bullet-\sigma$ relation
\citep{2005SSRv..116..523F}. Then, $\sigma_h^3 \propto M_{12}^{3/4}$ obtains. On the other hand, 
for a fixed galaxy mass, we
have $m_* \propto 1/N$ and therefore $T_{rlx} \propto M_{12}^{3/4} N/\rho$. Since in our N-body models, 
$\rho(r)$ and $n(E,t)$ are unchanged and only $\sigma$ changes as $M_{12}$ is varied, we find that 
the hardening rate scales with $M_{12}$ and $N$ as $s \propto  M_{12}^{-7/4} N^{-1}$. The case of a 
triaxial nucleus is different: $J$ for each star is not conserved, thus stars may precess into the 
loss cone on a time scale $T_{pr} \ll T_{rlx}$ \citep{2004ApJ...606.788P}; and $T_{pr}$ will
depend only on the global gravitational potential of the galaxy. In this case, the mass flux into
the loss cone $m_* \mathcal F_{lc}(E,t) \propto m_* n(E,t)/T_{pr}(E,t)$, and also $s(t)$, will be 
independent of the number $N$ of stars.

In Figure~\ref{fig1}, we see that $s(t)$ is N-dependent in a spherical nucleus, while it is
$N$-independent in the merging one. In the former case, $s(t) \propto N^{-\alpha}$, with $\alpha=0.45$
and $0.75$ for binaries of $M_{12}=0.005$ and $0.1$. These results can be interpreted as follows. 
In the empty loss cone limit ($\alpha=1$), the stars repopulate the loss cone at a rate $\propto 
T_{rlx}^{-1}$ much lower than the that with which they are ejected by the binary, which is $\propto 
T_{dyn}^{-1}$. In the full loss cone limit ($\alpha=0$), stars enter the loss cone at a rate which 
is similar to the rate at which they are ejected by the binary. A measure for the loss cone 
refilling rate is given by 
\citep{1977ApJ...211..244L}
\begin{equation}
q(E) \equiv \left( \frac{\delta J}{J_{lc}} \right)^2,
\label{eqn5}
\end{equation}
where $\delta J$ is the mean change in $J$, per orbital period, of a star on a low-$J$. 
\begin{figure}
\vspace{-0.4cm}
\epsscale{1.3}
\plotone{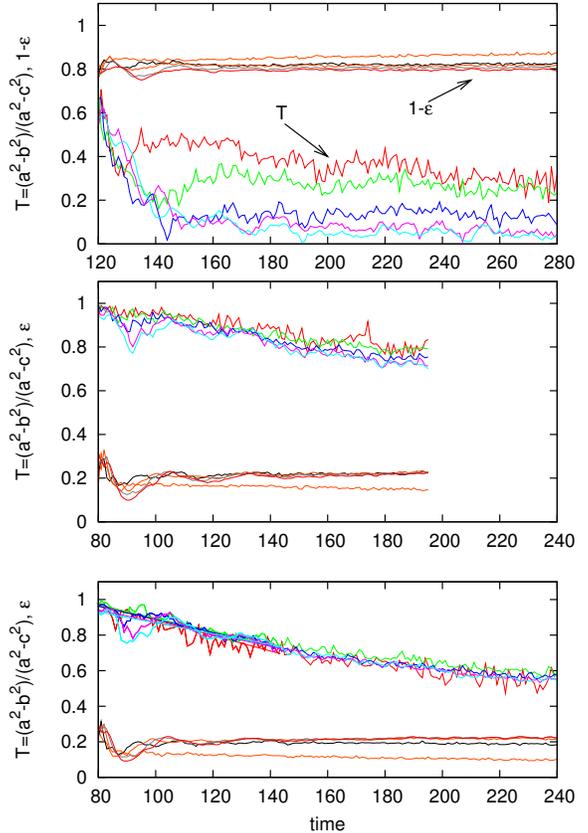}
\vspace{-1cm}
\caption{Triaxiality $T$ and mass flattening $\epsilon=(a-c)/a$ of merging nuclei. Shown are 
merging binaries of total mass $M_{12}=0.005$ with $L/L_c=0.6$ (upper panel), $M_{12}=0.01$ and 
$0.02$ with $L/L_c=0.14$ (almost radial mergers, in the middle and lower panels). $T$ and
$\epsilon$ are measured in five mass shells between $r=0$ and $r=2.5$, each of width $\Delta r=0.5$. 
Triaxiality decreases over time, the faster the heavier the binary is. Mass flattening is constant.}
\label{fig2}
\end{figure}
In the limit when $q(E) \ll 1$, the loss cone is said to be empty; while $q(E) \gg 1$ in the full 
loss cone limit. For a given nucleus, and for  $r>r_h$, we expect $\delta J$ to be independent of 
$M_{12}$. As a result, the weaker dependence of $\langle s \rangle$ on $N$ for lighter binaries, 
placed in a spherical nucleus, follows from $q \propto M_{12}^{-1/2}$; at 
the same $E$, $q(E)$ of the $M_{12}=0.005$ binary is $\sim 4.5$ larger than that of the $M_{12}=0.1$ 
one. We would need to use a larger $N$ for the lighter binaries, $\langle s \rangle \propto 
N^{-0.45}$, 
to enter deep into the empty loss cone limit $\langle s \rangle \propto N^{-1}$; the heavier 
binary, $\langle s \rangle \propto N^{-0.75}$, almost reaches this limit. The dependence 
of $\langle s \rangle$ on $M_{12}$ is more straightforward to interpret. For the spherical case, 
the lighter binary is expected to harden at a rate $\sim 20^{7/4}$ higher than the heavier,
which is indeed the case. In the merger case, $m_* \mathcal F_{lc}(E)$ is $N$-independent and 
therefore $\langle s \rangle \propto M_{12}^{-1}$. Since the mass ratio between the binaries is 
$2$, $\langle s \rangle$ also differs by a corresponding factor of two.  
%\vspace{1.cm}
\begin{figure}
\epsscale{1.3}
\vspace{-0.4cm}
\plotone{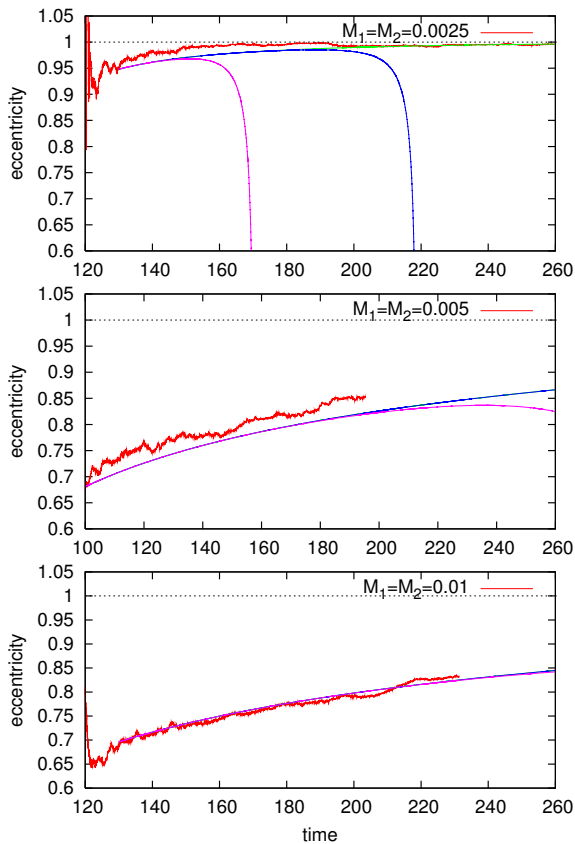}
\vspace{-1.cm}
\caption{Long term eccentricity evolution. Red and green lines represent
NB and semi-analytic evolution without radiation reaction. Blue and magenta 
lines correspond to semi-analytic solution, including radiation reaction, for 
$M_{12}=10^6 M_\odot$ and $M_{12}=10^8 M_\odot$, respectively.}
\label{fig3}
\end{figure}

Following \cite{2004ApJ...606.788P}, we measure the triaxiality of the nucleus with 
$T=(a^2-b^2)/(a^2-c^2)$. \footnote{Models with $T=0.25$ and $0.75$ correspond to moderately oblate 
and prolate shapes, respectively.} Figure~\ref{fig2} depicts the evolution of $T$ and of the 
flattening $\epsilon=1-c/a$ for several mass shells of merging nuclei. The value of 
$T$ of each remnant, immediately after the merger, depends on the initial $L/L_c$. In the case 
of a near radial merger, $L/L_c=0.14$, the remnant is prolate and evolves over time towards an 
oblate spheroidal shape; for $L/L_c=0.6$ the remnant is an oblate spheroid from the very beginning. 
The triaxiality decreases over time, and the rate at which it changes is faster the larger $M_{12}$ 
is. The triaxiality remains significant, in the inner mass shells, until the binary reaches the 
relativistic phase in all models with the smallest (and more realistic) values of $M_{12}$, and 
also in most of the other cases. The flattening $\epsilon \sim 0.2$ is constant throughout in 
all cases, so the asymptotic shape of the merger is that of an oblate spheroid. We conclude that 
the rather mild triaxiality created during the merger supports a family of centrophilic orbits 
that keep the loss cone full ($\alpha=0$) at all times until the binary reaches relativistic 
separations $\sim a_{GW}$. 

\section{Eccentricity evolution and time scales for coalescence}
\begin{figure}
\vspace{-0.4cm}
\epsscale{1.3}
\plotone{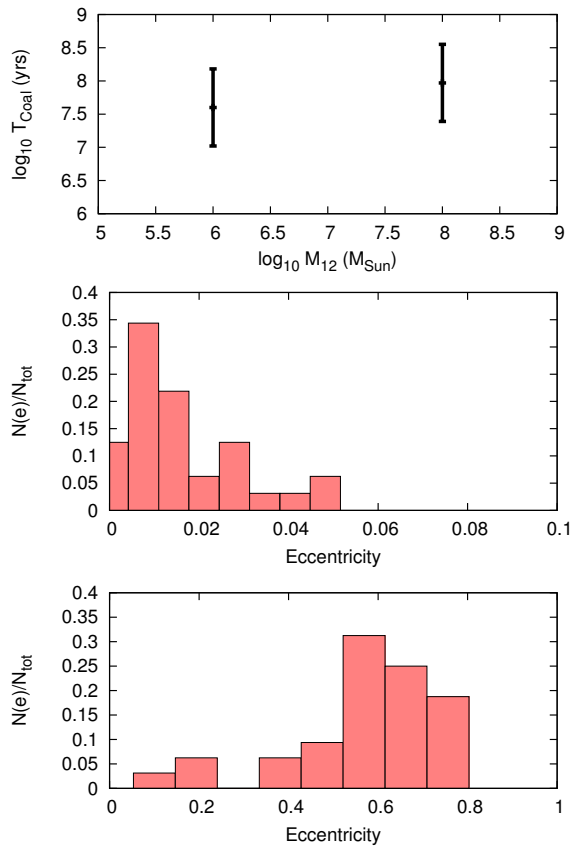}
\vspace{-1cm}
\caption{Upper panel: Range of coalescence time for binaries with $M_{12}=10^6 M_\odot$ and 
$10^8 M_\odot$. Middle panel: distribution of eccentricities, for $M_{12}=10^6 M_\odot$, at 
$a_{bin}=100 R_{Schw}$. Lower panel: distribution of eccentricities, for $M_{12}=10^8 M_\odot$, 
when $f_{orb}=2 f_{GW}=10^{-8} {\rm Hz}$ for PTAs.}
\label{fig4}
\end{figure}
%\vspace{-1cm}
The hardening rate, due to the slingshot ejection of stars, is found to be in our $N$-body 
simulations essentially independent of the binary's orbital elements, while that due to GWs 
is strongly dependent on them: $d/dt(1/a)_{GW} \sim |{\dot a}/a^2|_{GW} \propto a^{-5} (1-e^2)^{-7/2}$ 
\citep{1964PhRv..136.1224P}. As a result, the time a binary takes to coalesce depends strongly 
on its eccentricity. In paper I, we did follow this evolution self-consistently with N-body 
simulations of rotating King models. 
Since such calculations are extremely CPU-intensive, we estimate the full evolution using a 
semi-analytic approach \citep{1996NewA....1...35Q}. The advantage is that we can calibrate the 
average hardening rate $\langle s \rangle$ with our N-body simulations---which would remain a 
free parameter otherwise---, and thus make quantitative predictions on both the coalescence times 
and the long term eccentricity evolution. 

The evolution of the MBHB orbital elements, including the effect due to orbital energy 
lost to GWs, is given by
\begin{eqnarray}
\label{eqn6}
\frac{d}{dt} \left( \frac{1}{a}  \right) & = & \frac{d}{dt} \left( \frac{1}{a}  \right)_{st} + 
                                           \frac{d}{dt} \left( \frac{1}{a}  \right)_{GW}   \nonumber  \\
\frac{de}{dt} & = & \left( \frac{de}{dt} \right)_{st} + \left( \frac{de}{dt} \right)_{GW}. 
\end{eqnarray}
The GW terms are given in \cite{1964PhRv..136.1224P}. The eccentricity evolution, driven by 
the stars, is obtained from three-body scattering experiments
\begin{equation}
\label{eqn7}
\left( \frac{de}{dt} \right)_{st} = K(e) \ a \ \langle s \rangle,
\end{equation}
where $K(e)=e(1-e^2)^{k_0}(k_1+k_2e)$ and the constants are taken from \cite{1996NewA....1...35Q}. 
In order to assess the quality of the fits to the $N$-body results, Figure~\ref{fig3} compares
the $N$-body evolution of the binary with that obtained from the semi-analytic model. We take 
as initial conditions for the integration of equations~(\ref{eqn6}) an instant of time in the 
early hard binary phase. Given the differences between our $N$-body models and the assumptions 
embodied by the semi-analytic description the agreement is quite remarkable. 

We then include the GW terms due to radiation reaction to compute the time it takes for the binary 
to coalesce. To scale our models to binaries with $M_{12}=10^6 M_\odot$ and $M_{12}=10^8 M_\odot$, we 
adopt the most recent observational values for the mass normalization of the Milky Way nucleus, 
$M(<1 \rm{pc})=10^6 M_\odot$ \citep{2009A26A...502...91S} and use the $M_\bullet-\sigma$ relation 
to extrapolate to different MBH masses. The results are shown in the upper panel of Figure~\ref{fig4}. 
We see that coalescence times range between $T_{coal} \sim 10^7$ yrs and $\sim {\rm few} \times 
10^8$ yrs. These times are not longer that the mean time between successive major mergers. In 
contrast, for a spherical nucleus, coalescence times for the lower mass would become $\sim 
{\rm few} \times {\rm Gyr}$, while binaries with $\gtrsim 10^8 M_\odot$ would stall \citep{PB11}. 

We also follow the long term evolution of the eccentricity. In the $N$-body runs, the binaries 
become bound with high eccentricities (up to $e \sim 0.95$) on average---in agreement with 
previous works \citep{2009ApJ...695..455B,2009JPhCS.154a2049P}. Since LISA will be sensitive to 
the inspiral signal of $10^6 M_\odot$ binaries, it is important for data analysis purposes to 
estimate whether they will enter the band with non-negligible eccentricity ($e \gtrsim 10^{-4}$) 
\citep{2010arXiv1005.5296P}. The middle panel of Figure~\ref{fig4} displays the distribution of 
eccentricities at $a=100 R_{Schw}$\footnote{$R_{Schw}=2 G M_{12}/c^2$, where $c$ is the speed of 
light, is the Schwarzschild radius.}---most binaries will not be fully circularized by then. We 
expect therefore that the eccentricity in the LISA band will be non-negligible. 
Finally, the lower panel of Figure~\ref{fig4} depicts the eccentricity distribution at $f_{GW}=
2 f_{orb}=10^{-8} {\rm Hz}$ for the PTA band. We see that eccentricities are quite high---peaking 
at $e \sim 0.6$. The results presented here concerning the coalescence times and eccentricity 
growth corroborate recent three-body scattering studies---which had to treat the average hardening 
rate $\langle s \rangle$ as a free parameter \citep{2010ApJ...719..851S}.

\section{Summary}
With our results, we are moving closer towards a consistent solution to the {\it Final Parsec 
Problem}, and thus of providing a dynamical substantiation to the cosmological scenario where 
prompt coalescences are assumed during the course of galaxy evolution \citep{2007MNRAS.377.1711S,
2010A&ARv..18..279V}. Our results suggest that the formation of eccentric binaries, followed 
by a quick orbital decay, could result from the expected development of global non-axisymmetries 
in galaxies after they merge. Our gas-poor merger models show only rather mild departures from 
axisymmetry and a small amount of rotation; we believe that stronger departures from 
axisymmetry---to be expected from higher amount of rotation---, and the presence of gas will 
only reinforce our conclusions. It seems, therefore, probable that prompt coalescences result 
from mergers of irregular galaxies expected to be common at high redshift. Based on our prompt 
MBHB coalescence results, we expect that LISA will see $\sim 10-{\rm few} \times 10^2$ events 
per year depending on the MBH seed model \citep{2009CQGra..26i4033S,2010A&ARv..18..279V}.
Moreover, even though GWs circularizes the MBHBs during the late relativistic phase of 
inspiral, they are likely to have some residual ($e \gtrsim 0.001-0.01$) eccenticity when 
entering the LISA band and a broad distribution ($0.4 \lesssim e \lesssim 0.8$) in the PTA band.

\ \ \ \ \ 

\acknowledgements 
MP acknowledges support by DLR (Deutsches Zentrum f\"ur Luft- und Raumfahrt). 
We acknowledge support by the Chinese Academy of Sciences Visiting Professorship
for Senior International Scientists, Grant Number 2009S1-5 (The Silk Road Project)
(RS and PB). The special supercomputer Laohu at the High Performance
Computing Center at National Astronomical Observatories, funded by Ministry of
Finance under the grant ZDYZ2008-2, has been used. Simulations were also performed
on the GRACE supercomputer (grants I/80 041-043 and  I/84 678-680 of the
Volkswagen Foundation and 823.219-439/30 and /36 of the Ministry of Science,
Research and the Arts of Baden-W\"urttemberg). We thank the DEISA Consortium
(http://www.deisa.eu), cofunded through EU FP6 projects RI-508830 and RI-031513,
for support within the DEISA Extreme Computing Initiative. The Kolob cluster is
funded by the excellence funds of the University of Heidelberg in the Frontier
scheme.

%\clearpage


\begin{thebibliography}{32}
\expandafter\ifx\csname natexlab\endcsname\relax\def\natexlab#1{#1}\fi

\bibitem[{{Aarseth}(2003)}]{2003gnbs.book.....A}
  {Aarseth}, S. {Gravitational N-Body Simulations} (Cambridge, UK: Cambridge 
  University Press, November 2003.)

\bibitem[{Armitage \& Natarajan(2005)}]{2005ApJ...634..921A}
     {Armitage}, P. \& {Natarajan}, P. 2005, \apj, 634, 921

\bibitem[{{Begelman}, {Blandford} \& {Rees}(1980)}]{1980Natur.287..307B}
     {Begelman}, M.~C., {Blandford}, R.~D. \& {Rees}, M.~J. 1980, {\it Nature}, 287, 307

\bibitem[{{Berczik} {et~al.}(2005){Berczik}, {Merritt} \& {Spurzem}}]{2005ApJ...633..680B}
  {Berczik}, P., {Merritt}, D., \& {Spurzem}, R. 2005, \apj, 633, 680

\bibitem[{{Berczik} {et~al.}(2006){Berczik}, {Merritt}, {Spurzem} \&
  {Bischoff}}]{2006ApJ...642L..21B}
  {Berczik}, P., {Merritt}, D., {Spurzem}, R. \& {Bischoff}, H.~P. 2006, \apjl, 642, 21

\bibitem[{{Berczik} {et~al.}(2011){Berczik}, {Berentzen}, {Nitadori}, {Preto} \&
  {Spurzem}}]{PBNPS11}
  {Berczik}, P., {Berentzen}, I., {Nitadori}, K. {Preto}, M. 
  \& {Spurzem}, R. 2011, to be submitted to ApJ

\bibitem[{{Berczik} {et~al.}(2011){Berczik}, {Nitadori}, {Hamada} \&
  {Spurzem}}]{BNHS11}
  {Berczik}, P., {Nitadori}, K., {Hamada}, T. \& {Spurzem}, R.,
   2011, in preparation


\bibitem[{{Berentzen} {et~al.}(2009){Berentzen}, {Preto}, {Berczik},
  {Merritt} \& {Spurzem}}]{2009ApJ...695..455B}
  {Berentzen}, I., {Preto}, M., {Berczik}, P., {Merritt}, D. \& 
  {Spurzem}, R. 2009, \apj, 695, 455  (Paper I)

\bibitem[{{Callegari} {et~al.}(2011){Callegari}, {Kazantzidis}, {Mayer},
  {Colpi}, {Bellovary}, {Quinn} \& {Wadsley}}]{2011ApJ...729...85C}
    {{Callegari}, S., {Kazantzidis}, S., {Mayer}, L., {Colpi}, M., 
    {Bellovary}, J.~M., {Quinn}, T. \& {Wadsley}, J.}, 2011, \apj, 729, 85

\bibitem[{Colpi \& Dotti(2009)}]{2009arXiv0906.4339C}
{Colpi}, M. \& {Dotti}, M. 2009, {\it arXiv:0906.4339 }, 
Invited Review to appear on Advanced Science Letters (ASL), 
Special Issue on Computational Astrophysics, edited by Lucio Mayer

\bibitem[{{Dehnen}(1993)}]{1993MNRAS.265..250D}
{Dehnen}, W. 1993, \mnras, 265, 250

\bibitem[{Ferrarese \& Ford(2005)}]{2005SSRv..116..523F}
{Ferrarese}, L. \& {Ford}, H. 2005, {\it Space Sci. Rev.}, 116, 523

\bibitem[{{Harfst} {et~al.}(2007) {Harfst}, {Gualandris}, {Merritt}, {Spurzem}, 
	{Portegies Zwart} \& {Berczik}}]{2007NewA...12..357H}
        {Harfst}, S. and {Gualandris}, A. and {Merritt}, D. and {Spurzem}, R. and 
	{Portegies Zwart}, S. \& {Berczik}, P. 2000, {\it New Astronomy}, 12, 357

\bibitem[{Lightman \& Shapiro(1977)}]{1977ApJ...211..244L}
   Lightman, A.~P. \& Shapiro, S.~L. 1977, \apj, 211, 244

\bibitem[{{Khochfar \& Burkert}(2006)}]{2006A26A...445..403K} 
      {Kochfar}, S. \& {Burkert}, A. 2006, \aap, 445, 403

\bibitem[{{Merritt} \& {Poon}(2004)}]{2004ApJ...606.788P}
{Merritt}, D. \& {Poon}, M.Y. 2004, \apj, 606, 788

\bibitem[{Merritt {et~al.}(2007)}]{2007ApJ...671...53M}
      {Merritt}, D., {Mikkola}, S. \& {Szell}, A.  2007, {\apj}, 671, 53

\bibitem[{{Milosavljev\'ic} \& {Merritt}(2003)}]{2003ApJ...596..860M}
{Milosavljevi{\'c}}, M. \& {Merritt}, D. 2003, {\apj}, 596, 860

\bibitem[{{Quinlan}(1996)}]{1996NewA....1...35Q}
      {Quinlan}, G.~D. 1996, {\it New Astronomy}, 1, 35

\bibitem[{{Quinlan} \& {Hernquist}(1997)}]{1997NewA....2..533Q}
{Quinlan}, G.~D. \& {Hernquist}, L. 1997, {\it New Astronomy}, 2, 533

\bibitem[{{Perets} \& {Alexander}(2008)}]{2008ApJ...677.146P}
{Perets}, H. \& {Alexander}, T. 2008, \apj, 677, 146

\bibitem[{{Peters}(1964)}]{1964PhRv..136.1224P}
{Peters}, P.C. 1964, {\it Phys Rev}, 136, 1224

\bibitem[{{Porter} \& {Sesana}(2010)}]{2010arXiv1005.5296P}
{Porter}, E. \& {Sesana}, A. 2010, {\it arXiv:1005.5296} 

\bibitem[{{Preto} {et~al.}(2009)}]{2009JPhCS.154a2049P}
  {Preto}, M., {Berentzen}, I., {Berczik}, P., {Merritt}, D. \& 
  {Spurzem}, R. 2009, {\it Journ of Phys Conf Ser}, 154, 012049

\bibitem[{{Preto} \& {Amaro-Seoane}(2010)}]{2010ApJ...708L..42P}
{Preto}, M. \& {Amaro-Seoane}, P. 2010, \apjl, 708, 42

\bibitem[{{Preto} {et~al.}(2011){Preto}, {Berentzen},{Berczik},
  \& {Spurzem}}]{PB11}
  {Preto}, M., {Berentzen}, I., {Berczik}, P. \& 
  {Spurzem}, R. 2011, to be submitted to \apj

\bibitem[{{Sch{\"o}del} {et~al.}(2009){Sch{\"o}del}, {Merritt}, \&
  {Eckart}}]{2009A26A...502...91S}
{Sch{\"o}del}, R., {Merritt}, D., \& {Eckart}, A. 2009, \aap, 502, 91

\bibitem[{{Sesana}(2010)}]{2010ApJ...719..851S}
{Sesana}, A. 2010, \apj, 719, 851

\bibitem[{{Sesana} {et~al.}(2007){Sesana}, {Volonteri} 
        \& {Haardt}}]{2007MNRAS.377.1711S}
{Sesana}, A.,{Volonteri}, M. \& {Haardt}, F.  2010, \mnras, 377, 1711

\bibitem[{{Sesana} {et~al.}(2009){Sesana}, {Volonteri} 
        \& {Haardt}}]{2009CQGra..26i4033S}
{Sesana}, A.,{Volonteri}, M. \& {Haardt}, F.  2009, 
                    {\it Classical and Quantum Gravity}, 26, 4033

\bibitem[{{Spitzer}(1987)}]{1987degc.book.....S}
{Spitzer}, L. 1987, {Dynamical evolution of globular clusters} (Princeton, NJ,
  Princeton University Press, 1987, 191 p.)

\bibitem[{{Tremaine} {et~al.}(1994){Tremaine}, Richstone, Byun, Dressler,
  Faber, Grillmair, Kormendy, \& Lauer}]{1994AJ....107..634T}
{Tremaine}, S., Richstone, D.~O., Byun, Y.-I., Dressler, A., Faber, S.~M.,
  Grillmair, C., Kormendy, J., \& Lauer, T.~R. 1994, \aj, 107, 634

\bibitem[{{Volonteri}(2010)}]{2010A&ARv..18..279V}
{Volonteri}, M. 2010, \aapr, 18, 279

\bibitem[{{Yu}(2002)}]{2002MNRAS.331..935Y}
        {Yu}, Q. 2002, \mnras, 331, 935



\end{thebibliography}
\end{document}